\documentclass{article}
\usepackage{spconf,amsmath,graphicx,hyperref}

\usepackage{booktabs,multirow,siunitx}
\usepackage[table]{xcolor}
\definecolor{toplinegray}{gray}{0.45}
\newcommand{\topgray}[1]{\textcolor{toplinegray}{#1}}

\title{Voices as Handles: Reasoning about Speaker Identity with Frozen Text LLMs}

\name{Runqiu Xu, Zhisheng Zheng, David Harwath}
\address{The University of Texas at Austin, Austin, TX, USA}

\begin{document}
\ninept
\maketitle

\begin{abstract}
Multi-user voice agents must track who said what across dialogue sessions. Text LLMs are attractive backbones for such agents, but transcripts alone do not expose acoustic speaker identity, leaving the model without a persistent reference for linking information to speakers across sessions. We address this gap by introducing \textbf{Speaker Handles}, soft-token representations that expose acoustic speaker identity to a frozen text LLM for cross-session speaker-dependent reasoning. A three-stage curriculum trains a lightweight projector, with fewer than 0.1\% of the backbone's parameters, to map speaker embeddings into these handles. Establishing whether the resulting handles truly support cross-session speaker-dependent reasoning is challenging with existing benchmarks because textual cues can partially reveal fact ownership. We therefore present \textbf{SpeakerBind}, a controlled shared-agent benchmark in which overlapping facts across users require correct cross-session speaker attribution. Speaker Handles achieve 97.40--98.36\% accuracy on VoxCeleb1 and 70.40\% on SpeakerBind, close to the 71.88\% topline. These results show that the proposed Speaker Handles provide an efficient way to integrate acoustic speaker identity into frozen text LLMs for speaker--content reasoning.

\end{abstract}
\begin{keywords}
Speaker-attributed reasoning, speaker representations, large language models, multi-speaker spoken language understanding, benchmarking
\end{keywords}

\vspace{-8pt}
\section{Introduction}
\vspace{-8pt}

Voice agents that interact with multiple users need to link utterances to specific user identities and maintain these associations across speaker turns and sessions. For a voice agent shared by multiple users, answering ``What is on my schedule for tomorrow?'' requires recognizing the speaker's identity and then linking that identity to previous utterances made by the same user. Explicit textual speaker labels can provide these associations to a Large Language Model (LLM), but require consistent speaker identities to be assigned upstream of reasoning. We instead ask whether acoustic speaker identity can be exposed directly to a frozen text LLM as a reusable reference, allowing it to recover speaker relations across utterances and sessions without textual speaker IDs.

General audio-language models such as Voxtral~\cite{liu2025voxtral} and Qwen3.5-Omni~\cite{qwenteam2026qwen35omnitechnicalreport} integrate acoustic perception with language generation, yet speaker--content binding remains challenging even for recent omni-modal systems. Speaker-focused systems address specific aspects of multi-speaker processing through diarization, speaker-aware recognition, or dedicated acoustic conditioning~\cite{wang2024diarizationlm,speakerlm,dixtral}. These approaches rely on specialized model training, speaker-specific preprocessing, or both. In practice, however, developers may lack the resources to train or adapt a special-purpose speech LLM and instead need to rely on an existing text LLM as the reasoning model. With transcripts supplied separately, this leaves the question of how acoustic speaker identity can be exposed to the frozen LLM as a reusable reference across utterances.

Continuous prompts and lightweight cross-modal connectors have been used to expose speaking style and speaker characteristics to frozen or lightly adapted LLMs ~\cite{li2021prefix,li2023blip2,kang2025paralinguistic,thebaud2025enhancing,thebaud2026verification,speakerllm}, but these settings do not require speaker representations to serve as reusable references across utterances. In cross-session dialogue reasoning, the model must match recurring speakers and bind them to conversational facts.

To provide these reusable references, we introduce latent \emph{Speaker Handles}, soft-token representations projected from utterance-level speaker embeddings and supplied alongside transcripts in place of textual speaker labels. A lightweight projector produces these handles, while both the speaker encoder and text LLM remain frozen. The projector is trained with a three-stage curriculum.
Each utterance is encoded independently. The LLM uses the resulting handles to match speakers across occurrences and reason over their associated statements, without an upstream assignment of consistent textual speaker IDs.

Existing benchmarks evaluate speaker-attributed understanding~\cite{m3slu,sun2026msubench}, participants' information access~\cite{fantom}, and target-speaker meeting QA~\cite{dixtral}. However, our analysis finds that even after explicit textual speaker labels are removed, dialogue transcripts can retain cues that reveal speaker identity or answer-relevant structure, consistent with prior studies~\cite{wu2024justasr,lee2026hear}. Consequently, strong performance on these benchmarks alone does not establish that the model is using the supplied speaker--content bindings.

To address this limitation, we introduce \emph{SpeakerBind}, a controlled shared-agent benchmark designed so that correct answers require binding facts to the appropriate speakers across sessions. In this benchmark, independently generated user--agent sessions are interleaved into a shared history. Different users discuss overlapping topics and predicates while providing different facts, so topical relevance alone does not determine fact ownership. To test whether performance actually depends on these speaker--content bindings, we further apply three controlled interventions: reassigning answer-relevant history to incorrect speakers, collapsing all turns to a single speaker, and assigning a distinct speaker identity to every turn. These interventions corrupt speaker--content bindings, thereby testing whether the handles function as persistent speaker references.

Empirically, Speaker Handles approach the performance of ground-truth textual speaker labels, while incorrect speaker assignments cause performance to collapse. Our contributions are threefold: (1) a method that casts speaker embeddings as reusable latent handles, enabling a frozen text LLM to reason about who spoke what; (2) a three-stage training curriculum for handle grounding, interface generalization, and structured identity--content binding; and (3) SpeakerBind, a controlled shared-agent benchmark, paired with binding interventions, for evaluating speaker-dependent reasoning across sessions. 

\begin{figure*}[t]
    \centering
    \includegraphics[width=\textwidth]{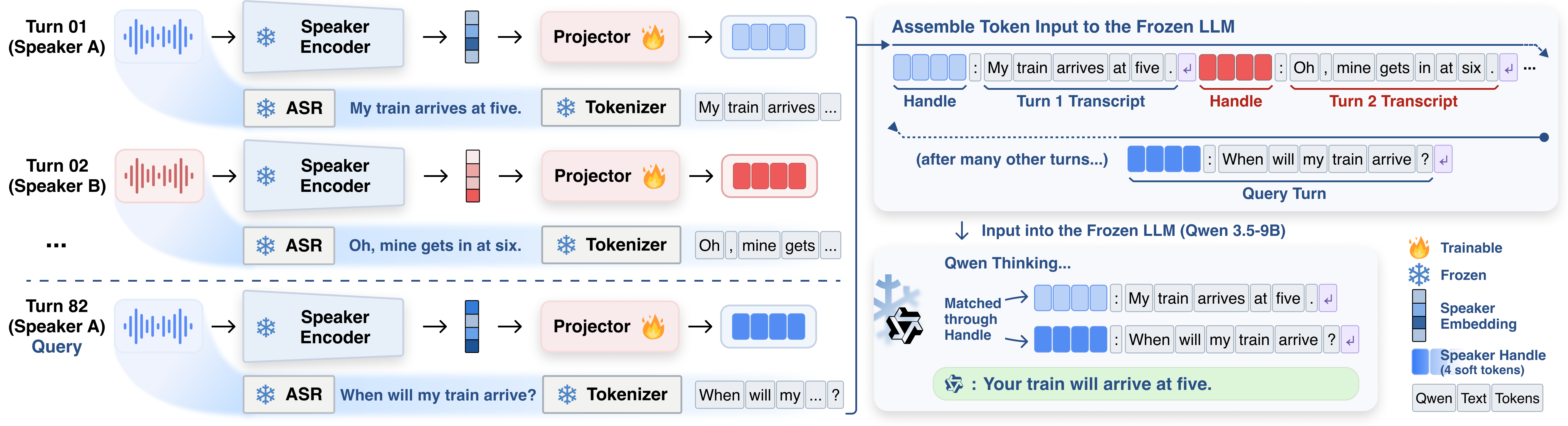}
    \vspace{-8mm}
    \caption{
        Speaker Handles as reusable speaker references for multi-speaker reasoning with a frozen text LLM.
    }
    \label{fig:framework}
\end{figure*}

\section{Latent Speaker Handles for Frozen LLMs}
\vspace{-6pt}
\label{sec:latent_speaker_handles}
We consider an ASR--LLM cascade for voice interaction, in which each spoken utterance is transcribed and the transcript is provided to a frozen text LLM. To expose speaker identity within this interface, we provide a Speaker Handle, a sequence of soft tokens projected from an utterance-level speaker embedding, alongside the corresponding transcript. We first describe the projector used to construct Speaker Handles and then the curriculum used to train it.

\vspace{-8pt}
\subsection{Speaker Projector}
\vspace{-4pt}

To construct a handle, a frozen speaker encoder first extracts an embedding \(\mathbf{e}_i=E(a_i)\) from an utterance \(a_i\). A trainable projector \(P_\theta\) maps this embedding to \(K\) vectors in the language model's input space,
\(
    \mathbf{H}_i=P_\theta(\mathbf{e}_i)
    =[\mathbf{h}_{i,1},\ldots,\mathbf{h}_{i,K}],
\)
where each \(\mathbf{h}_{i,k}\) has the language model's embedding dimension \(d\). We call \(\mathbf{H}_i\) a \emph{Speaker Handle}. It replaces the textual speaker label preceding an utterance and can also appear as a query-speaker reference, a roster entry, or a candidate in the prompt.

The handle is not a predefined name or discrete global speaker ID. When handles from multiple utterances are presented together, the language model can recover recurring speaker structure, group turns by identity, and reason over their associated content. These bindings can then be carried through subsequent reasoning. Figure~\ref{fig:framework} illustrates the inference-time interface. Each utterance is independently transcribed and encoded for speaker identity. The projected speaker handle is interleaved with the corresponding transcript tokens, producing the token sequence supplied to the frozen LLM. When a speaker recurs, the LLM can match the corresponding handles across turns and use this identity relation to retrieve and reason over that speaker’s earlier content.

\vspace{-6pt}
\subsection{Learning Speaker Handles}
\vspace{-4pt}

A projection into the input space does not by itself specify how a frozen language model should interpret the resulting vectors. We therefore design and train the projector through a three-stage curriculum as illustrated in Table~\ref{tab:handle_training_tasks}. Each input sample is a synthetic dialogue-like sequence and does not necessarily form a coherent conversation. We construct it by pairing independently sampled utterance texts with speech-derived speaker handles and concatenating the resulting turns. In task prompts, a handle may serve as a query-speaker reference or anchor. Some tasks add a temporary roster that maps participant names to independently derived handles.

\begin{table}[ht]
    \centering
    \caption{Training inputs and representative tasks.}
    \label{tab:handle_training_tasks}
    \scriptsize
    \setlength{\tabcolsep}{3pt}

    \begin{tabular*}{\columnwidth}{
        @{\extracolsep{\fill}}
        p{0.20\columnwidth}
        p{0.76\columnwidth}
        @{}
    }
        \toprule

        \multicolumn{2}{@{}l}{\textbf{Illustrative input}} \\
        \multicolumn{2}{@{}p{\columnwidth}@{}}{
            Turn 1 [\(\mathbf H_A^{(1)}\)]: ``The train was late.''} \\
        \multicolumn{2}{@{}p{\columnwidth}@{}}{  
            Turn 2 [\(\mathbf H_B^{(1)}\)]: ``We waited outside.''} \\
        \multicolumn{2}{@{}p{\columnwidth}@{}}{
            Turn 3 [\(\mathbf H_A^{(2)}\)]: ``It arrived after noon.''
        } \\

        \multicolumn{2}{@{}p{\columnwidth}@{}}{
            \textbf{Optional roster} [R]:\quad
            Ava \(=\mathbf H_A^{(r)}\),\;
            Ben \(=\mathbf H_B^{(r)}\).
        } \\

        \midrule
        \textbf{Task family} & \textbf{Representative training task} \\
        \midrule

        \multicolumn{2}{@{}l}{\textbf{Stage 1: Handle grounding}} \\[2pt]

        Retrieval \& copy &
        ``Copy exactly what \(\mathbf H_q\) said.'' \\[2pt]

        Anchor matching &
        ``Which turns share the speaker of Turn 2?'' \\[2pt]

        Verification &
        ``Does \(\mathbf H_q\) match the speaker of Turn 2?'' \\[2pt]

        \multicolumn{2}{@{}l}{\textbf{Stage 2: Interface generalization}} \\[2pt]

        Interface variants &
        Apply Stage 1 operations with bare handles, multiple candidates, varied handle positions, and alternative prompt formulations. \\[2pt]

        \multicolumn{2}{@{}l}{\textbf{Stage 3: Structured reasoning}} \\[2pt]

        Partition &
        Group turns by speaker, assign canonical speaker indices, or repair an incorrect assignment. \\[2pt]

        Temporal &
        Find a speaker's first, last, previous, next, or \(n\)-th occurrence. \\[2pt]

        Counts \& sets &
        Count speakers or occurrences and compare speaker sets across turns. \\[2pt]

        Consistency &
        Determine whether a collection belongs to one speaker or identify an inconsistent item. \\[2pt]

        Roster [R] &
        Map handles, turns, and quotations to participant names, or select the handle belonging to a named participant. \\[2pt]

        Fact binding [R] &
        ``Which fact is associated with Ava?'' \\

        \bottomrule
    \end{tabular*}

    \vspace{2pt}
    \scriptsize\raggedright
        [R] indicates roster-based tasks. Superscripts index utterances, and subscripts index ground truth identity.

    \vspace{-4pt}
\end{table}

Stage 1 grounds handles through query- and anchor-conditioned retrieval, copying, and verification. Stage 2 reformulates these operations using bare handles, alternative prompt formulations, multi-candidate prompts, and varied handle positions. Stage 3 adds global partitioning, temporal and set operations, error detection, and entity--content binding while replaying earlier tasks. The speaker encoder and language model remain frozen, and only \(P_\theta\) is optimized with causal language-model cross-entropy over answer tokens, averaged within each example.

\vspace{-10pt}
\section{SpeakerBind: Cross-session Speaker-Content Binding Benchmark}
\label{sec:speaker_contingent_reasoning}
\vspace{-4pt}

\subsection{Residual Speaker Cues in Existing Benchmarks}

Even after explicit textual speaker labels are removed, dialogue transcripts may retain cues about speaker structure that a text LLM can exploit as shortcuts. To probe this effect, we ask GPT-5.6-terra to estimate the number of participants from the unlabeled transcript, with turn boundaries and order preserved. We randomly sample 20 dialogues from each of MELD~\cite{poria2019meld}, AMI~\cite{carletta2006ami}, NOTSOFAR-1~\cite{vinnikov2024notsofar}, AISHELL-4~\cite{fu2021aishell4}, Earnings-21~\cite{delrio2021earnings21}, and ICSI~\cite{janin2003icsi}, for 120 dialogues in total. The predicted count is exact for approximately 34\% of dialogues and is within one participant of the reference for 72\%. Participant counting does not directly test speaker attribution, but the results suggest that unlabeled dialogue retains nontrivial information about the underlying speaker configuration. Such transcript-level shortcuts can therefore confound evaluation of whether a model actually uses the supplied speaker--content bindings.

\begin{figure*}[t]
    \centering
    \includegraphics[width=\textwidth]{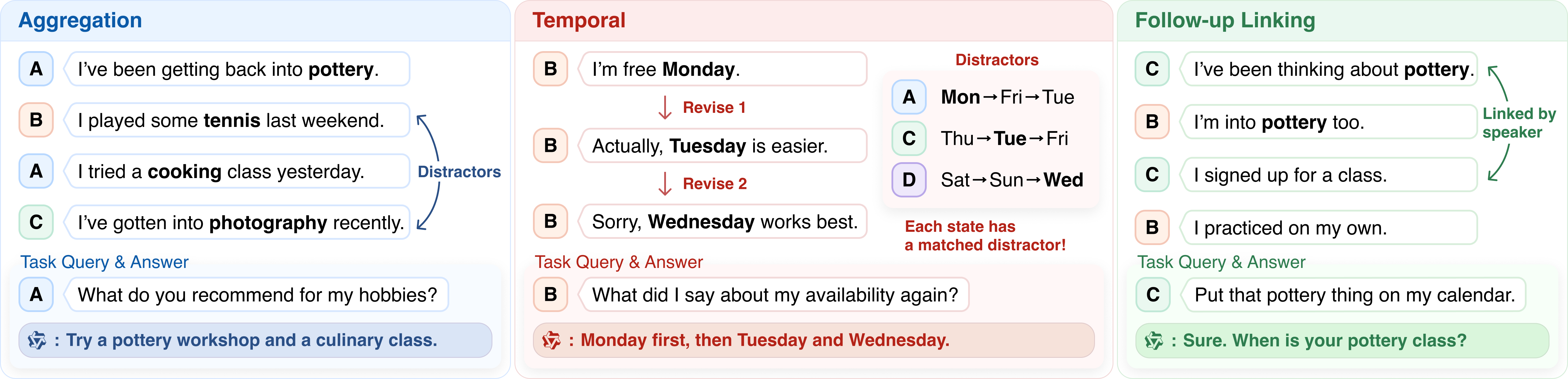}
    \vspace{-7mm}
    \caption{
        Conceptual examples of SpeakerBind, where each task requires reasoning over speaker–content bindings.
    }
    \label{fig:benchmark}
\end{figure*}

\subsection{Benchmark Construction}
\label{sec:benchmark_construction}
\vspace{-6pt}

To reduce these shortcuts, SpeakerBind uses a controlled shared-agent setting with multiple users interacting one-to-one with the same agent. These sessions are generated independently, then interleaved into a shared history. Users may discuss the same topics and predicates but report different facts, so topical relevance alone does not determine fact ownership. Questions therefore require the model to recover information associated with the query speaker across sessions while excluding competing facts from other users. Within this setting, we define three tasks that exercise this binding, as illustrated in Figure~\ref{fig:benchmark}.

\noindent\textbf{Aggregation.}
Each user reports multiple facts with the same topic and predicate across sessions. For example, a user reports practicing pottery and cooking in separate hobby sessions; other users report practicing different hobbies, while unrelated topics provide additional distractors. Answering \emph{What hobbies did I practice?} requires matching the speaker, topic, and predicate, collecting all relevant items, and excluding facts belonging to others.

\noindent\textbf{Temporal.}
Across successive sessions, users repeatedly update a state, such as availability. Each update states only the new value and does not repeat the previous one. Questions ask for the original, previous, or current value, or the entire trajectory. Distractors share portions of the target's trajectory. For a target chain \(\textbf{A} \rightarrow \textbf{B} \rightarrow \textbf{C}\), distractors follow \(\textbf{A} \rightarrow D \rightarrow E\), \(D \rightarrow \textbf{B} \rightarrow E\) and \(D \rightarrow E \rightarrow \textbf{C}\). A shared initial or final value therefore does not identify the target history. The model must track one speaker across sessions and preserve the order of updates.

\noindent\textbf{Follow-up Linking.}
Each user introduces multiple goals or interests (\emph{anchors}) and later reports a distinct follow-up for each. Anchor sessions introduce the goal without specifying or committing to a concrete follow-up. A user might discuss pottery and photography, then later report enrolling in a ceramics class and buying a camera. Another user may share the same interests but report different actions. A question specifies an anchor and asks what the query speaker subsequently did. The model must identify the correct speaker and link the specified anchor to its follow-up, while excluding the same user's other actions and other users' related actions.

\noindent\textbf{Dialogue generation.}
To obtain controlled overlap between users' histories, we generate dialogues from task-specific fact templates. Each template specifies a topic and predicate, together with a pool of candidate values. For example, a template for hobbies practiced can be instantiated with \emph{pottery}, yielding the fact \emph{I recently practiced pottery}. An LLM generates each user--agent dialogue independently. We then assign the resulting sessions to users and assemble a shared history, making cross-session ownership an explicit part of the construction. We derive reference answers from these assignments and the underlying facts and relations. Generation constraints and semantic checks aim to preserve the intended facts and prevent extra answers. Implementation details are given in Sec.~\ref{sec:setup}.

\vspace{-8pt}
\section{Experiments}
\vspace{-8pt}

\begin{table*}[t]
\vspace{-2mm}
\centering
\caption{
Results on existing benchmarks and SpeakerBind. The three interventions disrupt speaker structure in complementary ways: \textit{Reassigned} (\(^\dagger\)) changes the speaker assignments of answer-relevant historical turns while leaving the query speaker fixed; \textit{Same} (\(^\ddagger\)) maps all turns to one shared handle; \textit{Distinct} (\(^\S\)) assigns a unique handle to every turn, eliminating cross-turn identity recurrence. For SpeakerBind, expected original-gold accuracy under reassignment is 0.00, while task-specific chance levels for Agg./Temp./Follow-up are 0.00/37.05/50.00.
}
\label{tab:main_results}

\footnotesize
\setlength{\tabcolsep}{2.75pt}
\renewcommand{\arraystretch}{1.05}

\begin{tabular*}{\textwidth}
{@{\extracolsep{\fill}}
l
rrrrrrrr
rrrr
@{}}
\toprule

&
\multicolumn{3}{c}{VoxCeleb1} &
\multicolumn{1}{c}{M3-SLU} &
\multicolumn{2}{c}{FANToM} &
\multicolumn{2}{c}{NSF-QA} &
\multicolumn{4}{c}{SpeakerBind} \\

\cmidrule(lr){2-4}
\cmidrule(lr){5-5}
\cmidrule(lr){6-7}
\cmidrule(lr){8-9}
\cmidrule(lr){10-13}

Settings
& \multicolumn{1}{c}{O}
& \multicolumn{1}{c}{E}
& \multicolumn{1}{c}{H}
& \multicolumn{1}{c}{Acc.}
& \multicolumn{1}{c}{Fact QA}
& \multicolumn{1}{c}{ToM}
& \multicolumn{1}{c}{QA}
& \multicolumn{1}{c}{Sum. (/5)}
& \multicolumn{1}{c}{Agg.}
& \multicolumn{1}{c}{Temp.}
& \multicolumn{1}{c}{Follow-up}
& \multicolumn{1}{c}{Overall} \\

\midrule

\topgray{Ground-truth speaker labels \textit{(topline)}}
& \topgray{100.00}
& \topgray{100.00}
& \topgray{100.00}
& \topgray{65.25}
& \topgray{85.06}
& \topgray{65.71}
& \topgray{84.39}
& \topgray{3.04}
& \topgray{56.90}
& \topgray{80.05}
& \topgray{78.70}
& \topgray{71.88} \\

\textbf{Speaker Handles (Ours)}
& \textbf{98.36}
& \textbf{98.36}
& \textbf{97.40}
& \textbf{66.43}
& \textbf{78.51}
& \textbf{63.34}
& 83.69
& \textbf{2.83}
& 52.60
& \textbf{80.00}
& \textbf{78.60}
& \textbf{70.40} \\

Cascaded speaker identification
& \multicolumn{1}{c}{--}
& \multicolumn{1}{c}{--}
& \multicolumn{1}{c}{--}
& 60.05
& 71.38
& 57.18
& \textbf{84.16}
& 2.78
& \textbf{55.50}
& 77.95
& 76.40
& 69.95 \\

Transcript only \textit{(baseline)}
& \multicolumn{1}{c}{--}
& \multicolumn{1}{c}{--}
& \multicolumn{1}{c}{--}
& 53.41
& 52.18
& 47.39
& 81.76
& 2.28
& 0.20
& 25.43
& 47.00
& 24.21 \\

\midrule

Reassigned speaker bindings\(^\dagger\)
& \multicolumn{1}{c}{--}
& \multicolumn{1}{c}{--}
& \multicolumn{1}{c}{--}
& 51.41
& 42.41
& 48.85
& 66.80
& 1.90
& 0.00
& 0.25
& 8.60
& 2.95 \\

Same speaker for all turns\(^\ddagger\)
& \multicolumn{1}{c}{--}
& \multicolumn{1}{c}{--}
& \multicolumn{1}{c}{--}
& 52.40
& 55.17
& 52.63
& 78.08
& 2.09
& 0.10
& 35.88
& 47.30
& 27.76 \\

Distinct speaker for all turns\(^\S\)
& \multicolumn{1}{c}{--}
& \multicolumn{1}{c}{--}
& \multicolumn{1}{c}{--}
& 50.11
& 36.90
& 47.51
& 61.41
& 2.18
& 0.20
& 21.98
& 43.10
& 21.76 \\

\bottomrule
\end{tabular*}


\vspace{-4pt}
\end{table*}

\subsection{Setup}
\label{sec:setup}

All experiments use a frozen Qwen3.5-9B~\cite{qwen35} and a frozen VoxBlink2 ResNet293-LMFT speaker encoder~\cite{lin2024voxblink2}. The projector is a two-layer MLP with dimensions \(256 \rightarrow 512 \rightarrow (4 \times 4096)\), totaling 8.5M parameters. We use GELU activation and unit-normalize each output token. During training, we pair VoxCeleb2 speech~\cite{chung2018voxceleb2} with independently sampled LibriSpeech transcripts~\cite{panayotov2015librispeech}. For each example, we sample a speaker-to-turn assignment, use different recordings for repeated speakers, and derive the answer deterministically. Entity tasks add randomized names, aliases, and facts. The three stages run for 12K, 16K, and 4K steps, respectively. We use AdamW~\cite{loshchilov2019decoupled} with an effective batch size of 32 and a learning rate of \(10^{-5}\).

In evaluation, we compare Speaker Handles against several reference conditions and interventions. \emph{Ground-truth speaker labels} simulate access to a perfect upstream speaker-identification system and provide a textual-identity topline, while \emph{Transcript only} removes all explicit speaker-identity inputs. 
\emph{Cascaded speaker identification} serves as an explicit upstream baseline, where utterance-level speaker embeddings are first clustered into speaker identities and then provided to the LLM as textual speaker labels.
We also apply three interventions to the handle inputs. \emph{Reassigned speaker bindings} keep the query speaker fixed while assigning answer-relevant historical turns to other speakers. \emph{Same speaker for all turns} uses a single speaker identity throughout the history, whereas \emph{Distinct speaker for every turn} assigns a distinct, nonrecurring identity to each turn, eliminating between-speaker distinctions and cross-turn identity recurrence, respectively. Task-specific chance is computed analytically from the ambiguity induced by removing speaker identity, averaged over test examples.

Evaluation for existing benchmarks covers VoxCeleb1 verification~\cite{nagrani2017voxceleb,chung2018voxceleb2}, M3-SLU Task~2 same-person decisions in real dialogues~\cite{m3slu}, long-context FANToM factual and Theory-of-Mind (ToM) QA~\cite{fantom}, and speaker-related NSF-QA meeting QA and summarization~\cite{dixtral}. For NSF-QA, we use entity, topic, yes/no, detail, and target-speaker summary tasks. VoxCeleb1 uses its native audio, while M3-SLU and NSF-QA use original utterance audio. Since FANToM has no audio, each participant is mapped to a held-out VoxBlink2 speaker identity, with a different donor utterance sampled for each turn. These evaluation speakers are disjoint from projector-training speakers. GPT-5.6-terra judges FANToM Fact QA and NSF-QA open-ended answers for correctness against the ground truth and rates NSF-QA summaries from 1 to 5 by their agreement with the reference summaries.

SpeakerBind covers 149 task templates, including 54 for Aggregation, 61 for Follow-up Linking, and 34 for Temporal. These templates yield 1,540 session specifications (485, 327, and 728, respectively). Dialogues are generated from these specifications with SDialog~\cite{burdisso2026sdialog} using Qwen3.8-27B-FP8~\cite{qwen38}. We further apply LLM-based quality checks to verify that each session faithfully realizes its intended facts without introducing additional valid answers and to reject incompatible samples. The test set contains 1,000 histories per task and 6,000 questions in total. Aggregation and Follow-up Linking span 2--10 users and contribute one question per history, while Temporal spans 3--10 users and contributes four questions per history. Mean history length ranges from 138 to 187 turns across tasks. During evaluation, we assign speaker handles following the same procedure as for FANToM. Answers are judged by GPT-5.6-terra for correctness against the ground truth, following the same evaluation protocol described above.
\vspace{-8pt}

\subsection{Identity Preservation and Downstream Reasoning}
\label{sec:existing_benchmark_results}

Table~\ref{tab:main_results} summarizes results across speaker verification, existing downstream benchmarks, and SpeakerBind. For VoxCeleb1, we prompt the LLM to make speaker-verification
decisions on the standard test trials. Speaker Handles achieve 98.36\% accuracy on Vox1-O/E and 97.40\% on Vox1-H, showing that the frozen LLM can recover speaker relationships directly from the projected handles without first converting them into discrete speaker identities. Across existing downstream benchmarks, Speaker Handles consistently approaches the ground-truth textual-label topline, even slightly outperforming it on M3-SLU. It remains close on FANToM ToM and NSF-QA QA, while retaining strong performance on FANToM Fact QA and NSF-QA summarization. These results show that the projected handles preserve not only speaker-discriminative information but also much of the downstream utility provided by explicit textual speaker labels.

This trend is even clearer on SpeakerBind. Macro-averaged across Aggregation, Temporal, and Follow-up Linking, Speaker Handles achieve 70.40\%, compared with 71.88\% using ground-truth speaker labels. That is, the latent handle interface comes within 1.48 percentage points of the textual-label topline. The gap to ground-truth labels is only 0.10 points on Follow-up Linking, despite the need to resolve the query speaker and link an earlier anchor to its later follow-up. Since the benchmark templates and cross-session compositions are excluded from projector training, these results indicate that Speaker Handles transfer to unseen combinations of identity grounding and dialogue reasoning.

The intervention results further show that performance on SpeakerBind depends on the supplied speaker--content bindings and recurring speaker identity. Reassigning answer-relevant turns to incorrect speakers sharply degrades performance across all three tasks. Likewise, assigning the same speaker to all turns brings performance close to task-specific chance, while assigning a distinct speaker to every turn substantially degrades performance by eliminating cross-turn identity recurrence. In contrast, these interventions leave considerably more residual performance on several existing benchmarks, consistent with textual speaker cues, answer priors, or other content-based shortcuts. Together with the near-topline performance under correct bindings, these results provide evidence that the frozen LLM uses Speaker Handles as reusable speaker references for cross-session dialogue reasoning.

\vspace{-6pt}

\subsection{Ablation Study}

We ablate the training curriculum to examine the contribution of each stage and the effect of staged training. Table~\ref{tab:ablation} shows that the staged curriculum is important for learning effective Speaker Handles. Adding Stage 2 substantially improves speaker verification and Aggregation, while the full curriculum further improves Temporal reasoning without sacrificing overall downstream performance. In contrast, training all tasks at once sharply degrades downstream performance despite retaining substantial speaker-verification accuracy, indicating that speaker separability alone is insufficient for making the representations usable by the frozen LLM.
\vspace{-4pt}

\begin{table}[h]
    \centering
    \vspace{-10pt}
    \caption{Ablation of the staged training curriculum across speaker verification and downstream reasoning tasks.}
    \label{tab:ablation}
    \footnotesize
    \setlength{\tabcolsep}{3pt}
    \begin{tabular*}{\columnwidth}{@{\extracolsep{\fill}}lcccccc@{}}
        \toprule
        & & \multicolumn{3}{c}{SpeakerBind}
            & \multicolumn{2}{c}{NSF-QA} \\
        \cmidrule(lr){3-5}
        \cmidrule(l){6-7}
        Condition
        & Vox1-H
        & Agg.
        & Temp.
        & Follow-up
        & QA
        & Sum. \\
        \midrule

        Stage 1 only
        & 93.14
        & 47.20
        & 73.58
        & 79.00
        & \textbf{84.54}
        & 2.79 \\

        Stage 1 + 2
        & 97.24
        & 52.40
        & 75.15
        & \textbf{79.80}
        & 82.98
        & \textbf{2.83} \\

        All at once
        & 91.09
        & 7.00
        & 40.85
        & 66.20
        & 56.25
        & 2.67 \\

        \midrule

        Full (Ours)
        & \textbf{97.40}
        & \textbf{52.60}
        & \textbf{80.00}
        & 78.60
        & 83.69
        & \textbf{2.83} \\

        \bottomrule
    \end{tabular*}

    \vspace{-20pt}
\end{table}

\section{Conclusion}

\vspace{-6pt}

We introduced a lightweight projector that maps utterance-level speaker embeddings into latent Speaker Handles usable by a frozen text LLM. A staged curriculum teaches the projected handles to support identity operations and content binding. We also introduced SpeakerBind, a controlled benchmark for cross-session speaker–content binding across Aggregation, Temporal, and Follow-up Linking. The handles approach the performance of textual speaker labels on existing benchmarks and SpeakerBind. Ablations show that the staged curriculum is critical for making speaker representations usable by the frozen LLM beyond speaker discrimination alone. Reassigned speaker bindings drive accuracy toward the counterfactual expectation, while the same-speaker control approaches task-specific chance; assigning a distinct speaker to every turn similarly causes a large degradation.

\newpage

\section{Compliance with Ethical Standards}
This study uses only pre-existing speech and dialogue datasets and synthetically generated data. No participants were recruited or interacted with, and no new human-subject data were collected for this study.

\section{Conflict of Interest}
The authors have no relevant financial or nonfinancial interests to disclose.

\bibliographystyle{IEEEbib_short}
\bibliography{refs}

\end{document}